\documentstyle[aps]{revtex} 

\begin{document}

\title{Statistical properties of Charmonium spectrum and a new mechanism
of J/$\psi$ suppression}

\author{ Jian-zhong Gu$^{1,3,4}$, Hong-shi Zong$^{1}$, 
   Yu-xin Liu$^{2,1,3,}$,  
     En-guang Zhao$^{1,3}$ \\
\normalsize{$^{1}$ Institute of Theoretical Physics, Academia Sinica,
                         P. O. Box 2735,} \\
\normalsize{Beijing 100080, China}\\
\normalsize{$^{2}$ Department of Physics, Peking University, 
Beijing 100871,  China}\\
\normalsize{$^{3}$ Center of Theoretical Nuclear Physics, National 
             Laboratory of Heavy Ion Accelerator,}\\ 
\normalsize{Lanzhou 730000, China}\\
\normalsize{$^{4}$ Max-Planck-Institut fuer Kernphysik, Postfach 103980,
        D-69029, Heidelberg, Germany}}

\maketitle
\bigskip
\bigskip
\abstract{The statistical properties of Charmonium energy spectrum
determined by the Bethe-Salpeter equation  are investigated.
It is found that the regular motion of the $c\bar{c}$ system can be 
expected at a small value of color screening mass but the chaotic motion 
at a large one. It is shown that the level mixing due to color screening
serves as a new mechanism resulting in the J/$\psi$ suppression. Moreover,
this kind of suppression can occur before the color screening mass reaches
its critical value for J/$\psi$ dissociation. It implies that a strong
J/$\psi$ suppression is possible in the absence of dissociation of
J/$\psi$.}

\bigskip

PACS number(s): 14.40.Gx, 11.10.St, 25.75.Dw, 12.38.Mh

\bigskip 

\newpage

Strongly interacting matter of sufficiently high density is predicted
to undergo a transition to a state of deconfined quarks and gluons.
Deconfinement occurs when color screening shields a given quark from
the binding potential of any other quarks or antiquarks. Bound states
of heavy quarks, such as the J/$\psi$ or the $\Upsilon$, 
whose radii are much smaller than those of the usual mesons and nucleons, 
can survive in a deconfined medium until the temperature or density
becomes so high that color screening prevents their tighter binding.
It was theoretically predicted that a suppression of J/$\psi$ or
$\psi^{'}$ production due to the dissociation can be found in
relativistic heavy ion collision, which can serve as a clear signature
for the formation of Quark-Gluon-Plasma[1].
Subsequently this suppression was observed by NA38 collaboration[2].
However, successive research pointed out that such suppression could 
also exist in hadronic matter, even though caused by a completely different
mechanism[3]. Recently, an anomalously strong $J/\psi$ suppression has
been observed by the NA50 collaboration[4] and there have been a number
of attempts(see for example Refs.[5-7]) to explain it. 
However, the mechanism of J/$\psi$ suppression is still a question of
debate.

It is known that, for a complex system, the statistical properties of
the quantum spectrum and the complexity of the eigenfunctions reveal
frequently new features of the system. Because of the complexity of
the interaction and the temperature and density effect, the heavy
$q\bar{q}$ systems in hot matter can be regarded as complex systems.
Nevertheless, the spectral statistical properties and the complexity of
eigenfunctions of heavy $q\bar{q}$ systems have not yet been discussed.

In this paper we study the spectral statistical properties and the 
complexity of eigenfunctions of the $c\bar{c}$ system. The influence of 
color screening on the statistical properties and the mixing of  
eigenfunctions which characterizes the complexity of eigenfunctions
will be examined. And an alternative interpretation of the 
strong J/$\psi$ suppression is proposed.

To investigate the statistical properties of the $c\bar{c}$ system a
reasonable energy spectrum is crucial. Usually the bound states of the
$c\bar{c}$ system were studied in the nonrelativistic formulation[8,9].
However, the motion of a quark and an antiquark in a meson should be
relativistic. As pointed out in Ref.[10], the kinetic energy of the 
$c \bar{ c}$ system is about $13\%$ of the total energy and the ratio 
of relativistic corrections to the quark mass will not decrease with 
the increase of quark mass. As a result the bound state equation for 
$J/\psi$ should be relativistic. Since the Bethe-Salpeter (BS) 
equation[11] is the unique effective relativistic equation of two-body 
bound states and is consistent with quantum field theory, we take the 
BS equation to determine the energy spectrum of charmonium in this paper.

It is known that the full bound state BS equation, written in the two-sided
notation, reads[12]
\begin{equation}
(\eta_{1}\rlap/P+\rlap/p-m_{1})\chi_{{}_{P}}(p)(\eta_{2}\rlap/P-
\rlap/p+m_{2})= \frac{i}{(2\pi)^{4}}\int~d^{4}p'V(p,p';P)\chi_{{}_{P}}(p')
\end{equation}
where $\eta_{i}=\frac{m_{i}}{m_{1}+m_{2}}$ (i=1,2), $\chi_{P}$
is the wave function for the quark-antiquark system with total four
momentum $P$, $p$ is the relative four momentum. $V$ is the
interaction kernel that acts on $\chi_{P}$ and formal products
$V\, \chi_{{}_{P}}(p')$ take the form
$V\, \chi_{{}_{P}}(p')=V_{s}\, \chi_{{}_{P}}(p') +\gamma_{\mu}\otimes
\gamma^{\mu} V_{v}\, \chi_{{}_{P}}(p')$, in which $V_{s}$ and $V_{v}$
are the scalar, vector potential respectively.

With the standard reduction and spin-independent treatment being
implemented, the three-dimensional spin-independent reduced Salpeter
equation can be written as[13],
\begin{equation}
(M-E_{1}-E_{2})\phi(\vec{p})=\int~\frac{d^{3}p'}{(2\pi)^{3}}
\sum_{i=s,v}~F_{i}^{si}(\vec{p},\vec{p'})V_{i}(|\vec{p}-\vec{p'}|)
\phi(\vec{p'})
\end{equation}
where $\phi(\vec{p})=\int \! dp^{0}\,\chi_{{}_{P}}(p^{0},\vec{p}) $ is
the three-dimensional equal-time BS wave function. M is the mass of the
$q\bar{q}$ bound state. $E_{i}= (\vec{p}^{2}+m_{i}^{2})^{\frac{1}{2}}$,
i=1,2 represent a quark and an antiquark respectively. The functions
$F_{v}^{si}$ and $F_{s}^{si}$ are the transformation coefficients of
the spin-independent vector, scalar potential respectively when the
four dimensional BS equation is reduced to the three dimensional
spin-independent BS equation. Their expressions have been given
explicitly in Ref.[13].

It is definite that the momentum dependence of the interaction is
treated exactly in Eq.(2). And Eq.(2) gives a well-defined eigenvalue
problem for the masses M of the $q\bar{q}$ bound states in the momentum
space. To solve Eq.(2), we take the color screening potential between
the quark and antiquark[8]. In momentum space the color screening 
potential is written as

In our calculation, the reduced BS equation which has been described 
well in Refs.[12,13] and a color screened quark-antiquark potential[13] 
are implemented. In a momentum space, the potential with the color screening  
can be written as 
\begin{equation}
V_{s}(|\vec{p}-\vec{p'}|)=\frac{\sigma}{\mu}\delta^{3}(
\vec{p}-\vec{p'})-\frac{\sigma}{\pi^{2}}\frac{1}{
[(\vec{p}-\vec{p'})^{2}+\mu^{2}]^{2}} \, ,
\end{equation}
and
\begin{equation}
V_{v}(|\vec{p}-\vec{p'}|)=-\frac{2}{3\pi^{2}}\frac{\alpha_{s}}
{[(\vec{p}-\vec{p'})^{2}+\mu^{2}]},
\end{equation}
where $V_s$, $V_v$ refer to the scalar, vector potential respectively, 
and $\sigma$ is the string tension, $\alpha_{s}$ the effective
coupling constant, and $\mu$ the Debye screening mass.
The parameters including the value of the $\mu$ at zero temperature 
(denoted as $\mu _0$ hereafter) are determined by a least square  
fitting to the experimental data of $1S$, $2S$ and $3S$ states of the
$c\bar{c}$ system.   
The fixed parameters are $\sigma$=0.22 GeV$^{2}$, $\mu_{0}$=0.06 GeV,  
$m_{c}$=1.474 GeV, $\alpha_s(c \bar{c})$ = 0.47,  
which are within the scope of customary 
usage. Some of the obtained spin-averaged energy (mass) [13] spectrum of the
$c\bar{c}$ bound states at zero temperature ($T=0$) 
are listed in Table I. From Table I, one can observe that the mass 
spectrum obtained at present is more consistent with the experiment 
than the previous ones[8,9].
With the parameters $\sigma$, $m_c$ and $\alpha_{s}(c \bar{c})$  
fixed above, we evaluate the mass spectrum of the $c\bar{c}$ bound 
states for different $\mu$, so that the temperature dependence 
of the spectrum can be investigated.   
The calculation indicates that the critical value of the screening 
mass for J/$\psi$ to be dissociated is about $\mu_{c}$=0.90 GeV.

As the spectrum has been determined, we study the spectral statistics. 
To this end we exploit the well established random matrix theory 
(RMT)[14,15].
The RMT was initially used to describe the statistical properties of complex 
nuclear spectra in the early of 1950's[14]. In the RMT, the matrix of 
the Hamiltonian of the system, which is sufficiently complex and (or) 
governed by a very complicated or even unknown dynamics, is replaced 
by a random matrix with well-defined symmetry properties. Any desired 
quantity of the system is determined by performing an average over an 
ensemble of the random matrices of which the elements are distributed 
according to a certain probability distribution. The distribution 
is usually taken to be a Gaussian with a parameter $d$,
\begin{equation}
P(H)\:\propto\:exp(-Tr\{HH^{+}\}/2d^{2}),
\end{equation}
and the corresponding random matrix ensemble is called Gaussian 
ensemble. The underlying space-time symmetries obeyed by the 
system impose some important restrictions on the admissible 
ensemble. If the Hamiltonian is time-reversal and rotational invariant, 
the Hamiltonian matrices can be chosen to be real symmetric.  
The corresponding ensemble is called the Gaussian orthogonal 
ensemble(GOE).    
If the Hamiltonian is not time-reversal invariant, irrespective of its 
behaviour under rotations, the Hamiltonian matrices are complex Hermitian. 
The corresponding ensemble is called the Gaussian unitary ensemble (GUE). 
If the system is time-reversal invariant but not invariant under rotations, 
and it has half-odd-integer total angular momentum, the matrices are 
quaternion real. The corresponding ensemble is called the Gaussian 
symplectic ensemble (GSE).

If the Hamiltonian possesses other invariances, the corresponding 
ensemble can be the Gaussian unitary ensemble (GUE),  or the Gaussian  
symplectic ensemble (GSE).
The RMT has been applied successfully in many areas of physics, such as 
nuclear physics[16], quantum field theory[17] and the physics of disordered 
systems[18]. The success of the RMT is rooted in the fact that there exist 
so-called universal quantities which do not depend on the details of the 
dynamics but only on the underlying symmetries. This enables one to 
separate generic features from properties that do depend on the details 
of dynamics. The universal quantities usually mean the nearest neighbor 
level spacing distribution $P(s)$ and the spectral rigidity 
$\Delta _3 (L)$ which are powerful to analyze the spectral statistical 
properties of complex systems.

In present case the $c\bar{c}$ system is time-revsersal and rotational 
invariant. It is then a member of the GOE and its spectrum should be analyzed 
by comparing with  the predictions of the GOE.  
To make comparison with the universal and dimenssionless results of the 
RMT, it is necessary to make a transformation or normalization on the
spectrum. This operation is called the unfolding. The procedure is as 
follows. For a spectrum $\{E_{i}\}$, one separates its smoothed average 
part from its fluctuating part at first.
then counts the number of the levels below $E$ so that one can define a 
staircase function $N(E)$ of the spectrum,  
$$N(E) = N_{av}(E) + N_{fluct}(E)  .  $$
The unfolded spectrum can be finally obtained with the mapping 
$$ \tilde{E}_i = N_{av}(E_i) \, . $$
This unfolded levels $\tilde{E}_i$ are obviously dimensionless 
and have a constant average spacing of one, but the actual spacings 
exhibit frequently strong fluctuations. The nearest neighbor level 
spacings are defined as  
$s_i =\tilde{E}_{i+1} - \tilde{E}_i.$
The distribution $P(s)$ is defined as that $P(s)\, ds$ is the 
probability that $s_i$ lies within the infinitesimal interval
$[s, s\!+\!ds]$.

It has been shown that the nearest neighbor level spacing distribution 
$P(s)$ measures the level repulsion (the tendency of levels to avoid 
clustering) and short-range correlations between levels,  
which mean that, for a given level, its repulsion to the neighbor 
levels is limited within several level spacings[18,19].  
For the GOE, the  distribution can be expressed explicitly as 
$ P(s)=\frac{\pi}{2}~s~\exp(-\frac{\pi~s^{2}}{4})$, which can be 
deduced from Eq.(5) [15]. 
With the Brody parameter $\omega$ in the Brody distribution 
$P_{\omega}(s) = (1+\omega ) \alpha s^{\omega} \exp(-\alpha s^{1 + 
\omega} ) \, ( \alpha = [ \Gamma (\frac{2+ \omega}{1+ \omega} ) 
] ^{1+\omega} , \Gamma(x)$ is the $\Gamma$-function), the transition 
from regularity to chaos can be measured quantitatively. It is 
evident that $\omega =1$ corresponds to the GOE distribution, 
while $\omega =0$ to the Poisson-type distribution. A value 
$0 < \omega < 1$ means an interplay between the regular and 
the chaotic. 

As to the spectral rigidity $\Delta_{3}(L)$, it is defined as
\begin{equation}
\Delta_{3}(L)=\langle {\mbox{\large min}}_{{}_{A,B}}
\frac{1}{L}\int^{L/2}_{-L/2} [N(x) - Ax - B]^{2}dx \rangle ,
\end{equation}
where $N(x)$ is the staircase function of a unfolded spectrum in the 
interval $[-L/2, x]$. The minimum is taken with respect to the parameters 
$A$ and $B$. The average denoted by $\langle \cdots \rangle$ is taken over
a suitable energy interval over x. Thus from this definition
$\Delta_{3}(L)$ is the local average least square deviation of the
staircase function $N(x)$ from the best fitting straight line.
For the GOE the expected value of $\Delta_{3}(L)$ can only be evaluated
numerically, but it approaches the value
$ \Delta_{3}(L)\cong~\frac{1}{\pi^{2}}~(lnL-0.0687) $
for large L.
It has also been shown that the spectral rigidity $\Delta _{3}(L)$ 
signifies the long-range correlations of quantum spectra[18,19] 
which make it possible that for a chaotic spectrum very small fluctuation of the staircase 
function around its average can be found in an interval of given 
length (the interval may cover dozens of level spacings). 

It has been generally accepted that if the spectral statistical 
properties of a quantum system approach to the GOE, the motion of 
the system is chaotic, and if the statistical properties appear as  
Poisson-types ($P(s)=\exp(-s)$ and $\Delta_{3}(L)=\frac{L}{15}$) ), 
the motion is regular[18]. 

To perform a meaningful RMT analysis one has to sort the spectrum in
symmetry sectors corresponding to the symmetry of the Hamiltonian.    
Since the symmetry invariant subspaces are orthogonal to one another, 
each such symmetry invariant subspace is characterized by a specific 
set of quantum numbers.  The RMT analysis is performed on sets of 
the eigenlevels having the same quantum numbers. In view of the above 
consideration we take the first 200 levels out of 300 s-wave (spin-weighted) 
eigenvalues which have the same quantum numbers (e.g., $l=0$). 
We have noticed that p-wave eigenvalues (spin-weighted) have the same 
statistical properties as those of the s-wave ones. 
The spacing distribution $P(s)$ and spectral rigidity $\Delta_{3}$
for several color screening masses $\mu$ are shown in Fig.1.
The figure shows obviously that, when $\mu$ is small, for instance
0.06 GeV, both the $P(s)$ and $\Delta _{3}$ display the Poisson-type 
distribution. 
  This indicates that the spectrum is regular and
the levels are uncorrelated. The motion of the $c\bar{c}$ system is
then regular. With the increase of $\mu$ (for example $\mu=0.20$ GeV),
the $P(s)$ and $\Delta_{3}$ depart from the Poisson-types.
Then the spectrum becomes less regular and the level correlation
and repulsion increase gradually. As $\mu$ is large enough, say
$\mu$=0.60 GeV, both the $P(s)$ and the $\Delta_{3}$ appear the
GOE behavior. This means that the spectrum becomes extreme irregular
(chaotic) and the levels are strongly correlated.
Strong level repulsion can thus be expected and the motion of the
$c\bar{c}$ system is chaotic.

Fitting to the Brody distribution, we get the Brody parameter 
$\omega = 0.22, 0.51, 0.89$ for $\mu = 0.06, 0.20, 0.60$ (GeV) 
respectively. 
This also indicates that a transition from regular to chaotic motion 
can be induced in the $c\bar{c}$ system by the increasing of the 
color screening and the chaotic motion takes place when the hot 
matter is of high temperature. 
Recalling the idea that the $P(s)$ and $\Delta_{3}$ measures the 
correlation among the levels[18,19], one can recognize that the 
correlation of the $c\bar{c}$ resonances can be manifested by the 
$P(s)$ and $\Delta_{3}$. 
The calculated result shows thus that the $c\bar{c}$ resonances 
are strongly correlated at a high temperature. This information 
may be useful for us to analyze the relevant experiments.

As a matter of fact, both the level correlation and level repulsion 
stem from level mixing [20] which characterizes the complexity of
eigenfunctions. We analyze then the influence of the color screening on
the level mixing. For this purpose, we take the s-wave eigenfunctions 
of the $c\bar{c}$ system at temperature zero $|k \rangle$ (J/$\psi$, 
$\psi^{'}$, $\psi^{\prime\prime}, \cdots$, $\mu$=0.06 GeV, 
equivalent to T=0) as a basis and expand the eigenfunctions at 
a finite temperature $|i\rangle$ in this basis as
\begin{equation}
|i\rangle =\sum_{k}c_{k}^{i}|k\rangle \, ,
\end{equation}
with $ \sum_{k}(c_{k}^{i})^{2}=1. $

In order to measure the level mixing and delocalization of eigenfunctions
at a finite temperature, we define the following probability 
\begin{equation}
P(c^{2})=\frac{{\Delta}N(c^{2})}{N_{t}~\delta}\, ,
\end{equation}
where $c^{2}$ is the probability for the eigenfunctions at zero 
temperature to appear.
${\Delta}N(c^{2})$ is the number having the probability between 
$c^{2}$ and $c^{2}+\delta$ with $\delta$ being a small interval of 
$c^{2}$. $N_{t}$ is the total number of the eigenstates involved 
(at present $N_{t}=200$).
For the J/$\psi$ state (the ground state) at a finite temperature,
the $P(c^{2}$) versus $c^{2}$ for a few values of $\mu$ are shown in 
Fig.~2. It is evident that, when $\mu$ is small, for example 0.10 GeV,
the components with nonzero probabilities are limited within the lowest
states. It means that the components with small probabilities prevail 
and the distribution width of $P(c^{2}$) is rather small. 
In other words, the level mixing is localized.
With the increase of $\mu$, the number of the components with nonzero 
probabilities goes up. This leads to the increase of $P(c^{2})$ in the 
region of large probabilities (see the case $\mu$=0.20 GeV).
As $\mu$  approachs to $\mu_{c}$, the components with nonzero probability 
emerge on a large scale and become rather random (as shown by the cases
$\mu$=0.60, 0.90 GeV). Then, a wide distribution of P($c^{2}$) appears 
and the level mixing is strong and delocalized.
Referring to Fig.1, one can find that the quantum chaotic motion occurs 
when a strong level mixing emerges. It can also be found that, for higher 
eigenstates at a finite temperature, the level mixing changes with $\mu$ 
in a similar way as J/$\psi$ state. Furthermore the strong level mixing 
appears for these eigenstates even if $\mu$ takes a smaller value.

The feature of the effect of the level mixing induced by the variance of
$\mu$ is that the probability for the J/$\psi$ state at zero temperature
to appear in the J/$\psi$ state at a finite temperature descends with the
increasing of $\mu$.
Present result shows that at $\mu$=0.10, 0.20, the probability is
about 90$\%$, 70$\%$, respectively. As $\mu$ increases, the probability
decreases distinctly. For instance, when $\mu$=0.60 GeV, the probability
goes down to 10$\%$. Referring to Eq.(7) one can know that the
neighboring levels (e.g. $\psi^{'}$, $\psi^{''}$...) at a finite
temperature will also contribute to the J/$\psi$ at zero temperature.
However, their production probability is much smaller than that of
J/$\psi$ at a finite temperature (e.g. in p-p collisions, the ratio of
$\psi^{'}$ to $J/\psi$ is about 1/7[21], and in Pb-Pb collision it is
predicted to be $4\%$[22]). Then their contribution to the $J/\psi$ at
zero temperature should be very small. 
One can hence easily recognize that the production of J/$\psi$ at zero
temperature, which means in fact the probability for the charmonium states
at a finite temperature to include the $J/\psi$ state at zero temperature,
will be suppressed with the growth of $\mu$ and a strong suppression can
be expected when $\mu$ is large enough.
On the other hand, It has been well known that, in the deconfinement
scheme, the suppression due to the dissociation of J/$\psi$ occurs
only if $\mu$ reaches or exceeds the critical value $\mu_{c}$ [1].
Present analysis indicates that, the suppression due to the level
mixing can take place before $\mu$ reaches the critical value for
the dissociation to happen.
In addition, we find that for p-wave energy spectrum of the $c\bar{c}$ 
system, the transition to the chaos and the suppression due to the level 
mixing occur too. Moreover, the similar behavior can be found in the 
$b\bar{b}$ system.

To apply the above discussion to an actual physical situation, one needs to
know the specific dependence of the Debye screening mass $\mu$ on the
temperature T. Even though several studies on the properties of QCD
at finite temperatures and densities have been accomplished (see for
example Refs.[23-27]) and the general feature of the temperature dependence
$\mu (T)$ has been well established, the explicit expression $\mu (T)$ has
not yet been determined uniquely because a temperature-dependent running
coupling constant is involved.
Then the density dependence of $\mu$ is not clear now.
Any way, it is certain that the $\mu(T)$ increases with $T$ increasing.
The above calculated result manifests that, as the statistical
properties of the mass spectrum of the $c \bar{c}$ system is taken into 
account, the critical value $\mu _c$ for the strong suppression to take 
place gets lowered obviously from that in the deconfinement scheme
(from 0.9 GeV to 0.6 GeV). Then the corresponding critical temperature
$T_c$ decreases too. Consequently, much stronger suppression than that
predicted in the deconfinement scheme can emerge in the same experimental
status. The level mixing, or the chaotic motion, due to the color
screening is then an alternative mechanism to induce the strong J/$\psi$
suppression.

In conclusion, in this letter we have investigated the level statistics 
of the Charmonium energy spectrum in the framework of the BS equation 
and found that, with the increasing of the color screening mass, the transition 
from regular to chaotic motion can be induced for the $c\bar{c}$ system. 
The level mixing due to the color screening is then an alternative mechanism 
resulting in the strong J/$\psi$ suppression, and furthermore this kind 
of suppression can occur before $\mu$ reaches its critical value for the  
dissociation of $c \bar{c}$. This implies that a strong J/$\psi$ suppression
is possible in the absence of dissociation of the J/$\psi$. 

\vspace*{5mm}
   
Helpful discussions with Prof. Y. Z. Zhuo and Prof. B. Liu  
are highly appreciated. 
This work is supported partly by the National Natural Science 
Foundation of China. 
One of the authors (J.G) thanks also the support of the China 
Postdoctoral Science Foundation.
Another author (Y.L) thanks the support of Peking University too.

\newpage

\begin{description}
\item{Table I.} The calculated spin-averaged mass spectra $M_{nl}$ of 
the $c\bar{c}$ bound states and the comparison with 
experimental data and previous calculations.
\end{description}

\begin{center}
\begin{tabular}{|c|c|c|c|c|c|}
\hline
\multicolumn{1}{|c|}
{}  & {nl} &{Exp. (GeV) } & \multicolumn{3}{|c|}{Cal. (GeV)} \\
\hline
{ } &{ } &{ } &{Ref.[8]}
&{Ref.[9]} &{Ours} \\
\hline
$c\bar{c}$ & $1S$ & $J/\psi$(3.068)    & 3.070 & 3.070 & 3.067 \\ \cline{2-6} 
    {}     & $2S$ & $\psi^{'}$(3.663)  & 3.698 & 3.686 & 3.663 \\ \cline{2-6}  
    {}     & $3S$ & $\psi^{''}$(4.025) & 4.170 & 4.081 & 4.019 \\ \cline{2-6}  
    {}     & $1P$ & $\chi_{c}$(3.525)  & 3.500 & 3.505 & 3.526 \\ \hline
\end{tabular}
\end{center}

\newpage

\noindent{\large\bf Figure Captions}

\vspace*{1cm}

\begin{description}
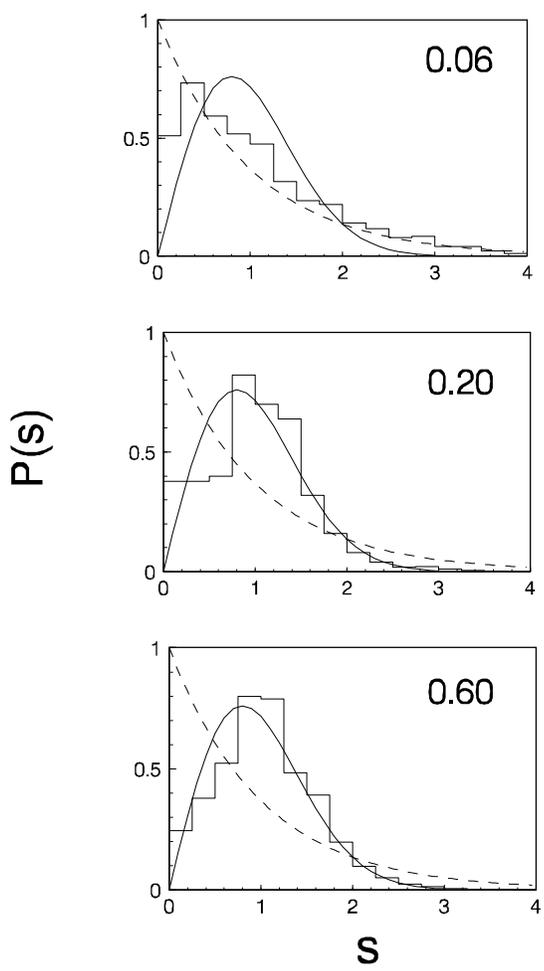
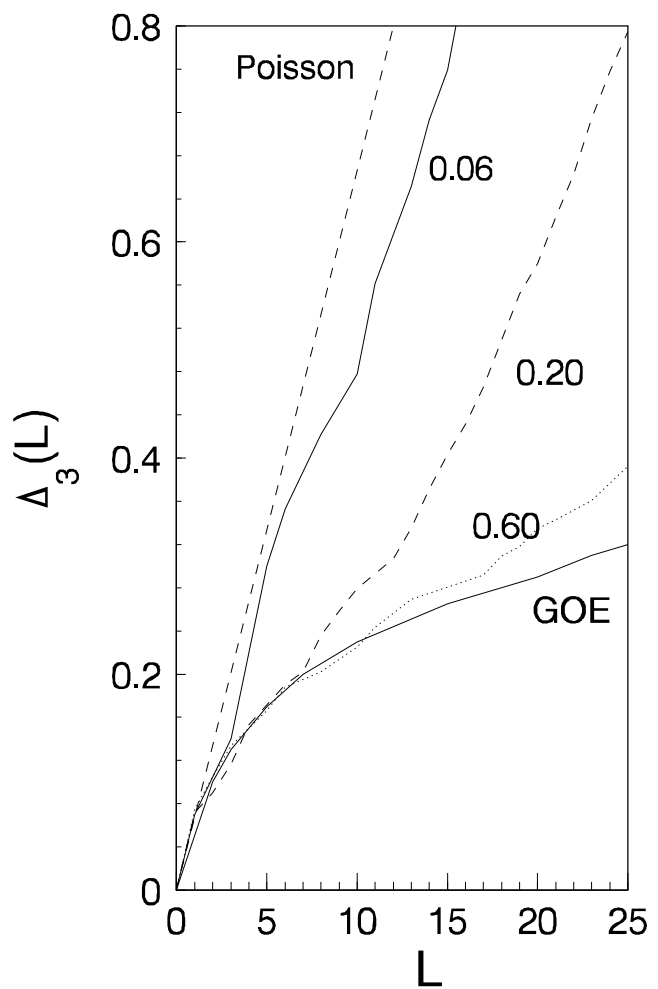
\item{Fig.1}
    The spacing distribution $P(s)$ and the spectral rigidity $\Delta_{3}$
of the s-wave Charmonium energy spectrum at different values of
color screening masses $\mu$ measured in GeV.
In (a) the dashed-line for Poisson distribution, the solid one for
the GOE distribution, and the histograms for our numerical results.

\vspace*{1cm}   

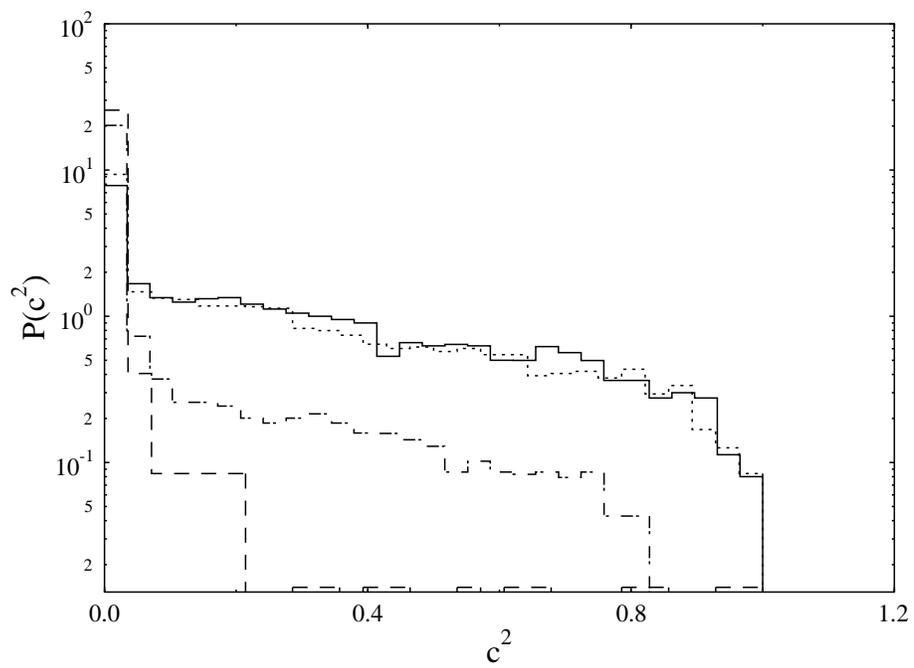
\item{Fig.2}
  The probability density $P(c^{2}$) of J/$\psi$ versus $c^{2}$
for various values of $\mu$.
The solid histogram for $\mu$=0.90 GeV, the dotted one for
$\mu$=0.60 GeV, the dash-dotted one for $\mu$=0.20 GeV and
the dashed one for $\mu$=0.10 GeV.
\end{description}

\end{document}